\documentclass[aps,preprint,notitlepage]{revtex4}
\usepackage{graphicx}
\usepackage{amsmath}
\usepackage{amssymb}
\usepackage{amsfonts}
\usepackage{bm}
\usepackage{color}
\usepackage[normalem]{ulem}
\newcommand{\be}{\begin{equation}}
\newcommand{\ee}{\end{equation}}
\newcommand{\ba}{\begin{eqnarray}}
\newcommand{\ea}{\end{eqnarray}}

\begin{document}

\title{Proposal of a Computational Approach for Simulating
Thermal Bosonic Fields in Phase Space}

\author{Alessandro Sergi}
\email{asergi@unime.it}
\affiliation{Universit\`a degli Studi di Messina,
Dipartimento di Scienze Matematiche e Informatiche,
Scienze Fisiche e Scienze della Terra,
viale F. Stagno d'Alcontres 31, 98166 Messina, Italy}

\affiliation{Istituto Nazionale di Fisica Nucleare, Sez. di Catania,
Catania 95123, Italy}

\affiliation{Institute of Systems Science, Durban University of Technology,
P. O. Box 1334, Durban 4000, South Africa}

\author{Roberto Grimaudo}
\email{roberto.grimaudo01@unipa.it}
\affiliation{Dipartimento di Fisica e Chimica dell’Universit\`a
di Palermo, Via Archirafi 36, I-90123 Palermo, Italy}

\author{Gabriel Hanna}
\email{gabriel.hanna@ualberta.ca}
\affiliation{Department of Chemistry, University of Alberta,
11227 Saskatchewan Drive Edmonton, Edmonton, AB~T6G~2G2, Canada}

\author{Antonino Messina}
\email{antonino.messina@unipa.it}
\affiliation{Dipartimento di Matematica e Informatica,
Universit\`a degli Studi di Palermo,
Via Archirafi 34, I-90123~Palermo, Italy}

\affiliation{Istituto Nazionale di Fisica Nucleare, Sez. di Catania,
Catania 95123, Italy}

\begin{abstract}
When a quantum field is in contact with a thermal bath, the vacuum
state of the field may be generalized to a thermal vacuum state,
which takes into account
the thermal noise. In thermo field dynamics,
this is realized by doubling the dimensionality of the Fock space of the system.
Interestingly, the representation of thermal noise by means of
an augmented space
is also found in a distinctly different approach based on the Wigner transform
of both the
field operators and density matrix, which we pursue here.
Specifically, the thermal noise is introduced by augmenting the
classical-like Wigner phase space by means of Nos\'e--Hoover chain thermostats,
which can be readily simulated on a computer.
In this paper, we illustrate how this may be achieved and discuss
how non-equilibrium quantum thermal distributions
of the field modes can be numerically simulated.\\
Published in {\it Physics} {\bf 2019}, 1(3), 402-411; {\bf https://doi.org/}10.3390/physics1030029 
\end{abstract}

\maketitle

\section{Introduction}
According to quantum field theory, the~vacuum is filled by fluctuating
quantum fields. Such~fields are present both in condensed matter
systems~\cite{sen,sen2,angew,ebbesen,ebbesen2,ebbesen3}
and, more generally, in~empty space-time, where their existence
is believed to be linked to dark matter and
dark energy~\cite{huang1,huang2,huang3,kerson-book,
dirac,dirac2,sudar1,sudar2,sudar3,eastman}.
Universal phenomena such as the emergence of order, symmetry breaking,
and phase transitions
are related to the {\it thermal degrees of freedom} 
the first time in order to stress that we are not using the expression
in a literal meaning.
of the fields~\cite{zinn-justin,bellac}.
Hence, the~study of the thermal excitation of the
quantum vacuum is of interest to many research areas~\cite{qv1,qv2,qv3,qv4,qv5}.
In thermo field dynamics~\cite{umezawa,umezawa2,das}, the~vacuum state
is generalized to a thermal vacuum state by doubling the dimension of the Fock space of the original vacuum. The~additional dimension
of the Fock space represent the degrees of freedom of the thermal bath, which 
are involved in the excitation and de-excitation processes
of the thermal~system.

In this paper, we illustrate an approach to simulate the
thermal distributions of bosonic fields~\mbox{\cite{tbf1,tbf2,tbf3,tbf4,tbf5,tbf6,tbf7,tbf8}} 
on a computer, which exploits the
Wigner formulation of quantum field theory~\cite{wff1,wff2,wff3,wff4,wff5,degroot}.
When considering a free field, it is useful to first express the
field operators and density matrices
in terms of creation and annihilation operators. 
The creation and annihilation operators are then mapped onto
canonical position and momentum operators,
which are finally Wigner transformed.
When the initial state of the field is pure,
the Wigner distribution function of the field is expressed as
an infinite product of the Wigner distribution functions~\cite{weinbub,ballentine,wigner}
of the single field modes. This~is also the case for a quantum thermal state of
an infinite set of free field modes, i.e.,~
the thermal Wigner function of the field is an infinite product
of the thermal Wigner functions of the modes.
In practice, it is possible to simulate
quantum thermal distributions with a different
temperature for each mode and to define a time-dependent
temperature. This is achieved by means of a technique called massive
Nos\'e--Hoover chain (NHC) thermostatting~\cite{nhc,massive-nhc},
which, in~our case, involves coupling a
separate NHC thermostat~\cite{nhc,massive-nhc} to each mode.
Since one requires a method that can be practically implemented on a computer,
the Fock space is truncated to become a many-dimensional (but finite)
Hilbert~space.

It should be noted that, in~spite of their theoretical differences, the~Wigner
representation of the thermal vacuum that is
obtained in our approach is analogous to the one given in
thermo field dynamics~\cite{umezawa,umezawa2,das}. 
As in thermo field dynamics, where the additional degrees of freedom 
represent the thermal bath, the~Wigner phase space of the
thermal system is augmented by the NHC degrees
of freedom~\cite{nhc,massive-nhc}.
In practice, because~each NHC can often be realized
with just two Nos\'e--Hoover thermostats chained to each other,
the dimension of the thermal Wigner phase space is~tripled.

A Wigner representation of the modes of the field~\cite{hillery} has certain
technical and conceptual advantages. First, since each mode is independently
thermostatted, it is easy to generate a non-equilibrium 
quantum thermal distribution of the field by assigning
each mode a different thermodynamic temperature.
Secondly, there is no technical difficulty in making the mode temperatures
time dependent. As~will be explained below,
this feature can be exploited to simulate the passage from a quantum
thermal distribution of the modes to a classical one.
The generalization of our formalism to situations where the field is
coupled to a spin system~\cite{swf1,swf2,swf3,swf4,swf5,swf6,swf7,swf8,swf9,swf10,swf11}
is left for future work.
Such~an endeavour could lead to new computational
experiments on spin dynamics, which to the best of our knowledge,
would be novel.
We propose one such computational experiment in the~Conclusions.

The paper is organized as follows. In~Section~\ref{sec:wbft},
we show how to represent the dynamics and averages of
bosonic field modes in Wigner phase space.
In Section~\ref{sec:ctwigner}, we discuss how a bosonic thermal state 
is described within the Wigner representation of quantum field theory.
We also illustrate how to control the temperature
of each field mode on a computer.
Finally, our conclusions and perspectives are given in Section~\ref{sec:concl}.

\section{Wigner Formulation of Bosonic Field~Theory}
\label{sec:wbft}

Let us consider a quantum bosonic free field, confined to
a finite region of space. This field possesses discrete mode frequencies,
$\omega_J$, and~bosonic creation and annihilation operators,
$\hat a_J$ and $\hat a_J^\dag$, where $J$ is an integer.
The Hamiltonian operator of the free field reads~\cite{wff4,hillery}:
\begin{equation}
\hat H_{(\hat a,\hat a^\dag)}
=\sum_J\hbar\omega_J\left(\hat a_J^\dag\hat a_J + \frac{1}{2}\right)\;,
\label{eq:Haadag}
\end{equation}

Let us now define the following canonical transformation:
\ba
\hat a_J &=&\frac{1}{\sqrt{2\hbar}}\left(
\lambda_J\hat Q_J+\frac{i\hat P_J}{\lambda_J}\right)
\;,\label{eq:aj}\\
\hat a_J^\dag &=&
\frac{1}{\sqrt{2\hbar}}\left(\lambda_J\hat Q_J-\frac{i\hat P_J}{\lambda_J}\right)
\;,\label{eq:ajdag}
\ea
where $\hat Q_J$ and $\hat P_J$ are Hermitian operators obeying the canonical
commutation relation $[\hat Q_J,\hat P_{J'}]=i\hbar\delta_{JJ'}$.
The scaling factor $\lambda_J$ is equal to $\sqrt{m_J\omega_J}$ when the field
consists of oscillators of mass $m_J$ or~is equal to
$\lambda_J=\sqrt{\hbar}(\omega_J/c)$ in the case of
an electromagnetic field~\cite{hillery}.
Upon substituting Equations~(\ref{eq:aj}) and~(\ref{eq:ajdag}) in
Equation~(\ref{eq:Haadag}), we obtain another representation for the
discretized quantum free field Hamiltonian,
\begin{equation}
\hat H_{(\hat Q,\hat P)}=\sum_J
\left[\frac{\omega_J\hat P_J^2}{2\lambda_J^2}
+ \frac{\omega_J\lambda_J^2\hat Q_J^2}{2}\right]
\label{eq:HQP}\;,
\end{equation}

Since any initial state of the field can be expanded in
the basis of Fock states, we consider a multimode Fock state.
This can be written as:
\ba
|\Phi\rangle &=& |n_1,n_2, ..., n_K, ...\rangle
=\frac{(\hat a_1^\dag)^{n_1}}{\sqrt{n_1+1}}
\frac{(\hat a_2^\dag)^{n_2}}{\sqrt{n_2+1}} \cdot \cdot \cdot
\frac{(\hat a_K^\dag)^{n_K}}{\sqrt{n_K+1}}\cdot\cdot\cdot|0\rangle
\nonumber\\
&=&\prod_{J=1}^\infty \frac{(\hat a_J^\dag)^{n_J}}{\sqrt{n_J+1}}|0\rangle
\equiv\prod_{J=1}^\infty|\Phi_J\rangle
\;,
\ea
where $n_J$ is the occupation number of state $J$.
To each state $|\Phi_J\rangle$, one can associate a single mode density matrix,
i.e.,~$\hat\rho_J=|\Phi_J\rangle\langle\Phi_J\vert$.
Therefore, the~total density matrix of the field is an
infinite tensor product of the single mode density matrices:
\be
\hat\rho=\prod_{J=1}^\infty\hat\rho_J\;.
\ee

We can now define the Wigner transform~\cite{weinbub,ballentine,wigner}
of each mode's density matrix as:
\be
W_J(Q_J,P_J)=\frac{1}{(2\pi\hbar)^d}\int dZe^{i P\cdot Z/\hbar}
\langle Q_J-\frac{Z}{2}\vert \hat\rho_J\vert Q_J+\frac{Z}{2}\rangle\;,
\ee
where $d$ is the dimension of the spatial region in which the field is confined.
The Wigner distribution function of the field is therefore given by:
\be
W(Q,P)=\prod_{J=1}^\infty W_J(Q_J,P_J) \;,
\ee
where the field-mode coordinates, $(Q,P)=(Q_1,Q_2,\ldots,P_1,P_2,\ldots$),
are c-numbers in Wigner phase space~\cite{ballentine,wigner}.

We now consider the quantum averages of generic bosonic field operators,
$\hat{\cal O}$, in~the second quantized~form:
\be
\langle\hat {\cal O}(t)\rangle={\rm Tr}(\hat\rho(t)\hat{\cal O})\;.
\ee
If $\hat{\cal O}$ is symmetrically ordered with respect to the
$\hat a_J$ and $\hat a_J^\dag$ operators~\cite{hillery}, that is,
\ba
\hat{\cal O}=\prod_{J=1}^\infty
\sum_{m,n=0}^\infty b_{nm}\frac{1}{2}\left[(\hat a_J^\dag)^n\hat a_J^m
+\hat a_J^m(\hat a_J^\dag)^n\right]
\equiv \prod_{J=1}^\infty\hat{\cal O}_J
\;,
\ea
where $b_{nm}$ are expansion coefficients,
then we can introduce the Wigner transform of $\hat{\cal O}$ as:
\be
{\cal O}_W(Q,P)=\prod_{J=1}^\infty\int dZ_J e^{iP_J\cdot Z_J/\hbar}
\langle Q_J-\frac{Z_J}{2}\vert \hat{\cal O}_J \vert Q_J+\frac{Z_J}{2}\rangle.
\ee
Under the restriction for the field observables mentioned
above, averages can then be calculated as~\cite{hillery}:
\be
\langle\hat {\cal O}(t)\rangle={\rm Tr}(\hat\rho(t)\hat{\cal O})
=\int dX W(Q,P,t) {\cal O}_{W}(Q,P)\;,
\ee
where, in~practice, in~order to obtain the Wigner transform of 
operators $\hat{\cal O}$ obeying the condition mentioned above,
one first uses the canonical transformations in Equations~(\ref{eq:aj})
and~(\ref{eq:ajdag}) and~then replaces the operators $(\hat Q_J,\hat P_J)$
with the c-numbers $(Q_J,P_J)$.
For example, the~Wigner transform of the field Hamiltonian
in Equation~(\ref{eq:Haadag}) reads: 
\be
H_{(Q,P)}=\sum_J
\left[\frac{P_J^2}{2\mu_J} + \frac{\mu_J\omega_J^2 Q_J^2}{2}\right]
=\sum_J H_W^J(Q_J,P_J)
\label{eq:HQPW}\;,
\ee
where $\mu_J=\lambda_J^2/\omega_J$. Equation~(\ref{eq:HQPW})
is just the c-number version of Equation~(\ref{eq:HQP}).

In the Wigner representation, the~quantum vacuum field equations
in the Heisenberg picture:
\begin{eqnarray}
\frac{d}{dt}\hat a_J(t)&=&\frac{i}{\hbar}
\left[\hat H_{(\hat a,\hat a^\dag)},\hat a_J(t)\right]
\label{eq:ddtaj}
\;,\\
\frac{d}{dt}\hat a_J^\dag(t)&=&\frac{i}{\hbar}
\left[\hat H_{(\hat a,\hat a^\dag)},\hat a_J^\dag(t)\right]
\label{eq:ddtajdag} \;,
\end{eqnarray}
become:
\begin{equation}
\frac{d}{dt} X_J = - H_X \overleftarrow{\nabla}\mbox{\boldmath$\cal B$}
\overrightarrow{\nabla} X_J(t)\;, \label{eq:Xclassdyna}
\end{equation}
where $X=(X_1,X_2,\ldots, X_J,\ldots)$ (with $X_J=(Q_J,P_J)$),
$\nabla =\partial / \partial X$, and~the direction of the arrow indicates the direction in which 
the operator acts. 
The matrix $\mbox{\boldmath$\cal B$}$ is the constant
symplectic matrix~\cite{goldstein}:
\be
\mbox{\boldmath$\cal B$}=\left[\begin{array}{cc}
 \mathbf{0} & \mathbf{1}\\ \mathbf{-1} & \mathbf{0}
\end{array}
\right] \;,
\ee
whose specific form gives rise to the Poisson bracket~\cite{goldstein} on the right hand side
of Equation~(\ref{eq:Xclassdyna}) {(where $\mathbf{1}$, for~example, denotes an infinite-dimensional block diagonal matrix of ones)}.
Equation~(\ref{eq:Xclassdyna}) is equivalent to Equations~(\ref{eq:ddtaj})
and~(\ref{eq:ddtajdag}), showing that the quantum dynamics of
a harmonic system may be exactly mapped onto a classical-like time evolution.
In this case, the~quantum character of the system enters
the description through the initial~conditions.

In the next sections, we will see how the use of the Wigner representation
allows 
one to exploit the many numerical algorithms 
originating from molecular dynamics simulation in order to propagate
the degrees of freedom represented in phase~space.
 
\section{Computer Simulation of Thermal Field~States}
\label{sec:ctwigner}

The field Hamiltonian in Equation~(\ref{eq:Haadag}) is isomorphic to a Hamiltonian of
a collection of non-interacting harmonic oscillators~\cite{qft,qft2,qft3,qft4}.
The thermal Wigner function, $W_J^{T}$, of~a harmonic mode has the
following analytical form~\cite{hillery}:
\be
W_J^{T}(Q_J,P_J)
=
\frac{\omega_J \tilde{\beta}_J(\omega_J)}{2\pi}
\exp\left[-\tilde{\beta}_J(\omega_J) H_W^J(Q_J,P_J)\right]
\label{eq:thermalW}\;,
\ee
where:
\be
\tilde{\beta}_J(\omega_J) = 
\frac{2 \tanh\left(\frac{\beta \omega_J}{2}\right)}{\omega_J},
\label{eq:tildebeta_J}
\ee
is a frequency-dependent inverse temperature with $\beta=1/k_{\rm B}T$,
and the Hamiltonian $H_W^J(Q_J,P_J)$ of the $J^{\rm th}$ mode is defined in 
Equation~(\ref{eq:HQPW}).
The thermal Wigner function of the field is:
\be
W^{T}(Q,P)=\prod_{J=1}^\infty W_J^T(Q_J,P_J)\;.
\label{eq:Wfieldthermal}
\ee

Equations~(\ref{eq:thermalW})--(\ref{eq:Wfieldthermal}) show that, in~the Wigner
representation, the~thermal state of a free field is given in terms
of an infinite product of single mode Wigner functions.
Each of these single mode Wigner functions is a Boltzmann factor
with a frequency-dependent temperature, as~defined
in Equation~(\ref{eq:tildebeta_J}).
Thus, the~thermal vacuum state requires each oscillator to have a minimum 
average energy consistent
with the temperature constraints fixed by Equation~(\ref{eq:tildebeta_J}).
In a numerical simulation, the~number of oscillators in Equation~(\ref{eq:Wfieldthermal}) for the thermal state of the free field must be truncated to a finite
number $N$, {i.e.},
\be
\tilde{W}^T(Q,P)=\prod_{J=1}^N W_J^T(Q_J,P_J)\;.
\label{eq:WNmodesthermal}
\ee

The dynamics of this field can be simulated by first sampling the initial $(Q_J,P_J)$ from the Boltzmann factors $W_J^T(Q_J,P_J)$
for each of the $N$ modes and then propagating the modes in time
by numerically solving Equation~(\ref{eq:Xclassdyna}).
Since Equation~(\ref{eq:Xclassdyna}) generates constant energy trajectories,
the~thermal equilibrium distribution in Equation~(\ref{eq:WNmodesthermal})
is an invariant of the energy conserving phase space flow of the free field. 
It should be noted, however, that the invariance is broken when
the field is coupled to another system.
The lack of invariance of the thermal distribution arises ultimately
from the non-conservation of the free field Hamiltonian in the presence
of the~coupling.

In addition to the case when the field is coupled to another system~\cite{swf1,swf2,swf3,swf4,swf5,swf6,swf7,swf8,swf9,swf10,swf11}
there are a number of cases when it is desirable to simulate
the dynamics of the thermal field state~\cite{tbf1,tbf2,tbf3,tbf4,tbf5,tbf6,tbf7,tbf8}
on a computer.
To this end, we employ a Nos\'e--Hoover chain (NHC)
thermostat~\cite{nhc,massive-nhc},
which may be theoretically defined in terms of
a quasi-Lie bracket~\cite{b1,quasi-Lie,quasi-Lie2,quasi-Lie3}.
In general, NHC thermostats are used to increase the ergodicity of
the dynamics of non-ergodic systems, such as a system of harmonic oscillators.
In NHC thermostatted dynamics, the~phase space point of oscillator $J$
is extended as follows:
\begin{equation}
\tilde{X}_J = (Q_J,\xi_1^J,\xi_2^J,\ldots,\xi_n^J,P_J,\chi_1^J,\chi_2^J,\ldots,\chi_n^J)\;,
\end{equation}
where $\xi_K^J$ and $\chi_K^J$ 
denote the position and momentum, respectively, of~
the $K^{\rm th}$ thermostat in the chain attached to mode $J$ and~
$n$ defines the length of the chain.
The energy of the NHC thermostat is given~by:
\be
H_{NHC}(\chi,\xi)=\sum_{J=1}^N\left[
\left(\frac{\left(\chi_1^J\right)^2}{2{\cal M}_{1}}
+Nk_{\rm B}T_J\xi_1^J\right)
+\sum_{K=2}^n \left(\frac{\left(\chi_K^J\right)^2}{2{\cal M}_{K}}
+k_{\rm B}T_J\xi_K^J\right)\right]
\;.
\ee

The temperature control of the physical coordinates, $X$, of~the field
by the fictitious NHC phase space coordinates, 
denoted collectively by $(\chi,\xi)$, is realized by solving a set of 
quasi-Hamiltonian equations 
of motion (shown below in Equation~(\ref{eq:Wclassdynanhc})).
The inertial parameters ${\cal M}_{K}$ (where \mbox{$K=1,\ldots,n$})
control the speed of the response of the thermostat variables to
the imbalance between the instantaneous kinetic energy of each mode,
$P_J^2/2\mu_J$, and~the kinetic energy corresponding to its desired temperature $T_J$, $k_{B}T_J$, where $k_{B}$ denotes the Boltzmann constant. Finally, 
the Hamiltonian of the field together with the NHC thermostat is: 
\be
H_{tot}=H_{(Q,P)} + H_{NHC}(\chi,\xi)
\;. \label{eq:HamFR}
\ee

In order to illustrate the theory, for~convenience,
we set the number of thermostats in the NHC to $n=2$. The~quasi-Hamiltonian
equations of motion are given by: 
\be
\frac{d}{dt}\tilde X_J(t) = - H_{tot} 
\overleftarrow{\nabla}\mbox{\boldmath$\cal B$}_{\rm NHC}
\overrightarrow{\nabla} \tilde X_J(t)
\;, \label{eq:Xclassdynanhc}
\ee
where $\nabla=(\partial/\partial Q_1,\ldots,\partial/\partial Q_J,\partial/\partial\xi_1^J,
\partial/\partial\xi_2^J,\partial/\partial P_J,\partial/\partial\chi_1^J,
\partial/\partial\chi_2^J,\ldots,\partial/\partial\chi_2^N)$ and: 
\be
\mbox{\boldmath$\cal B$}_{\rm NHC}
=\left[\begin{array}{cccccc}
\mathbf{0} & \mathbf{0} & \mathbf{0} & \mathbf{1} & \mathbf{0} & \mathbf{0} \\
\mathbf{0} & \mathbf{0} & \mathbf{0} & \mathbf{0} & \mathbf{1} & \mathbf{0} \\
\mathbf{0} & \mathbf{0} & \mathbf{0} & \mathbf{0} & \mathbf{0} & \mathbf{1} \\
\mathbf{-1} & \mathbf{0} & \mathbf{0} & \mathbf{0} & \mathbf{-P} & \mathbf{0} \\
\mathbf{0} & \mathbf{-1} & \mathbf{0} & \mathbf{P} & \mathbf{0} & \boldsymbol{-\chi_1} \\
\mathbf{0} & \mathbf{0} & \mathbf{-1} & \mathbf{0} & \boldsymbol{\chi_1} & \mathbf{0} \\
\end{array}\right]
\label{eq:BN}
\ee
is the antisymmetric matrix that generalizes the symplectic matrix,
defining the quasi-Hamiltonian NHC phase space flow ({where each element of the matrix
in Equation~(\ref{eq:BN}) is an $N\times N$
block diagonal matrix, e.g.,~$\mathbf{P}$ is an $N\times N$ block diagonal matrix containing the $P_J$'s along the diagonal).}
The right hand side of Equation~(\ref{eq:Xclassdynanhc}) defines
the quasi-Lie bracket~\cite{b1}, which generalizes the Poisson (Lie) bracket
to the case of NHC dynamics~\cite{nhc,massive-nhc}.
It should be noted that
the coupling between the field and the NHC thermostat does not
arise from the Hamiltonian in Equation~(\ref{eq:HamFR}),
but from the
the matrix $\mbox{\boldmath$\cal B$}_{\rm NHC}$ in the quasi-Hamiltonian equations of motion in Equation~(\ref{eq:BN}).
Finally, the~initial thermal Wigner function in the extended space is given by:
\be
\tilde {W}^T(\tilde X)=\prod_{J=1}^N W_J^T(Q_J,P_J)
\prod_{K=1}^2\delta\left(\xi_K^J-\xi_K^J(0)\right)
\delta\left(\chi_K^J-\chi_K^J(0)\right)
\;.
\label{eq:WNHC}
\ee

Its equation of motion reads~\cite{wignernose}:
\be
\frac{\partial}{\partial t}\tilde W^{T}(\tilde X,t) = 
H_{tot} 
\overleftarrow{\nabla}\mbox{\boldmath$\cal B$}_{\rm NHC}
\overrightarrow{\nabla} \tilde W^{T}(\tilde X,t)
-\kappa \tilde W^{T}(\tilde X,t)
\;, \label{eq:Wclassdynanhc}
\ee
where: 
\ba
\kappa=\sum_{K=1}^{\bf 6N}\frac{\partial\dot{\tilde{X}}_K}{\partial\tilde X_K}
=\sum_{K=1}^{6N}\sum_{\bf J=1}^{\bf 6 N}\frac{ \partial
{\cal B}_{\rm NHC}^{ \rm K J}}
{\partial X_K} \frac{\partial H_{\rm tot}}{ \partial X_J}
\ea
is the compressibility of the phase space.
{The emergence of the compressibility of the phase space
in Equation~(\ref{eq:Wclassdynanhc}) is a signature of the effects of the
quasi-Hamiltonian evolution of the coordinates on the Wigner function.
To understand fully how the compressibility arises, one can introduce
the Wigner--Liouville operator $i{\cal L}=- H_{tot}
\overleftarrow{\nabla}\mbox{\boldmath$\cal B$}_{\rm NHC}
\overrightarrow{\nabla}$ from Equation~(\ref{eq:Xclassdynanhc}) into the expression for the 
Wigner phase space average of a generic operator $\hat{\cal O}$ in the Heisenberg picture, i.e.,
\be
\langle {\cal O}(t)\rangle
=\int dX W^{\rm T}(X)e^{i\overrightarrow{\cal L}t}{\cal O}_{\rm W}(X)
\;.
\ee
Upon integrating by parts, one obtains the corresponding expression in the Schr\"odinger picture:
\be
\langle {\cal O}(t)\rangle
=\int dX W^{\rm T}(X)e^{-i\overleftarrow{\cal L}^\dag t}{\cal O}_{\rm W}(X)
\;,
\ee
where the adjoint Wigner--Liouville operator is given by:
\be
-i{\cal L}^\dag= H_{tot}
\overleftarrow{\nabla}\mbox{\boldmath$\cal B$}_{\rm NHC}
\overrightarrow{\nabla} -\kappa
\;.
\ee
}

From a very different perspective, Equations~(\ref{eq:WNHC}) and~(\ref{eq:Wclassdynanhc})
achieve a similar goal to that of
thermo field dynamics: 
the thermal state is represented in a manner analogous to a pure state,
but in a space with additional dimensions, {viz.,}
the extended Fock space in thermo field dynamics or the extended 
Wigner phase space in the current formulation.
The main difference is that the additional coordinates of the NHC thermostats
in the extended Wigner phase space are treated classically.

The possibility of separately controlling the temperature of each mode
by means of an NHC thermostat may lead to the design of computational
experiments
that, to~our knowledge, have not been performed before for quantum processes.
For example, starting from the quantum initial condition in Equation~(\ref{eq:WNHC}) 
and then setting $\tilde{\beta}_J=\beta$ for $J=1,...,N$ (which corresponds to
the classical limit) in the NHC dynamics generate a
nonequilibrium situation, from~which the field would
ultimately reach its classical thermal state.
In a situation where the field is no longer isolated,
it would be interesting to investigate how the time dependent classical limit
of the field would modify the dynamics of the coupled system.
Moreover, the~classical limit may be applied to groups of modes so that
one can study how each group affects the dynamics of the coupled~system.

\section{Conclusions}
\label{sec:concl}

In this paper, we discussed how pure and thermal states of free bosonic
fields~\cite{tbf1,tbf2,tbf3,tbf4,tbf5,tbf6,tbf7,tbf8} 
can be represented theoretically and simulated computationally.
The simulation protocol, which involves an NHC thermostat~\cite{nhc,massive-nhc}
coupled
to each field mode, propagates the dynamics of a thermal state living
in an extended Wigner space~\cite{wignernose}.
This is conceptually similar to what is done in thermo field dynamics~\cite{umezawa,umezawa2,das}.

For simplicity, the~theory was only applied to the free field case~\cite{wff4}, 
in order to demonstrate the technicalities associated with coupling
a different NHC thermostat~\cite{nhc,massive-nhc}
to each field mode.
However, this theory is intended for situations where the
field is coupled to a spin system~\cite{swf1,swf2,swf3,swf4,swf5,swf6,swf7,swf8,swf9,swf10,swf11}. In~this case, the~use of NHC thermostats
in the dynamics of the thermal state of the field would make it possible
to simulate processes that, to~our knowledge, have not been
investigated so far. For~example, 
NHC dynamics could allow one to simulate the transition from a vacuum field
state to a thermal vacuum state (or to thermally excited states of the field).
Then, one could study changes in the spin system's transport
properties in response to various field thermal processes. Such 
applications are left for future~work.

\vspace{6pt} 

\begin{flushleft}
{\bf Author Contributions:} The authors contributed equally to this~work.
\end{flushleft}

\begin{flushleft}
{\bf Funding:} This research received no external~funding.
\end{flushleft}

\begin{flushleft}
{\bf Conflict of interests:} The authors declare no conflict of~interest.
\end{flushleft}

\noindent
{\bf Abbreviations}
\noindent
The following abbreviations are used in this manuscript:\\

\noindent 
\begin{tabular}{@{}ll}
NHC & Nos\'e--Hoover chain 
\end{tabular}


\end{document}